# A rapid decrease in the rotation rate of comet 41P/Tuttle-Giacobini-Kresák

Dennis Bodewits[1], Tony L. Farnham[1], Michael S. P. Kelley[1], and Matthew M. Knight[1]

**Cometary outgassing can produce torques that change the spin state of the nucleus, influencing the evolution and lifetimes of comets (1,2). If these torques spin up the rotation to the point that centripetal forces exceed the material strength of the nucleus, the comet may fragment (3). Torques that slow down the rotation can cause the spin state to become unstable, but if the torques persist, the nucleus may eventually reorient itself and start to spin up again (4). Simulations predict that most comets will go through a short phase of changing spin states, after which changes occur gradually over long times (5). We report on observations of comet 41P/Tuttle-Giacobini-Kresák during its highly favourable close approach to Earth (0.142 au on April 1, 2017) that reveal a dramatic spin-down. Between March and May 2017, the nucleus' apparent rotation period increased from 20 hours to over 46 hours, reflecting a rate of change more than an order of magnitude larger than has ever been measured before. This phenomenon must be caused by a fortuitous alignment of the comet's gas emission in such a way as to produce an anomalously strong torque that is slowing the nucleus' spin rate. The behaviour of 41P suggests that it is in a distinct evolutionary state and that its rotation may be approaching the point of instability.**

The highly favourable apparition of comet 41P/Tuttle-Giacobini-Kresák (hereafter 41P) made it the target of observations worldwide for several months. We report on results from our observations of comet 41P obtained in March 2017 using the Large Monolithic Imager on the Lowell Observatory's 4.3-m Discovery Channel Telescope (DCT), and in May using the UltraViolet-Optical Telescope (UVOT) on board the Earth-orbiting Swift Gamma Ray Burst Mission (6; Extended Data Table 1).

We used comet-specific narrowband filters (7) on the DCT to capture the emission of cyanogen (CN) gas. CN coma structures have been used to infer rotational properties of otherwise unobservable comet nuclei since their discovery in comet 1P/Halley (8). Volatile ices at or near a comet's surface sublimate when exposed to sunlight during a comet's diurnal cycle. As the gas moves outwards, it and/or daughter species produced by photodissociation trace spirals or arcs that are diagnostic of the comet's rotation. CN is one of the most effective gases in this respect, owing to its large fluorescence efficiency in sunlight. Its use is widespread (9), and its connection to the rotation of comet nuclei has been verified by in situ missions like EPOXI (10). During our first epoch of observations, we identified rotating spiral arms, of which one is persistent while a second is visible for part of the rotation only (Fig. 1). The repetition of these features indicated a rotation period between 19.75 and 20.05 hours during March 5 – 9 (11).

For our second epoch, we adopted a photometric technique, using variations in the comet's brightness to measure periodicity. Although these techniques are measuring different characteristics, they both identify repetitions in their respective phenomena and we assume that the associated periodicities reflect the rotation of the nucleus. We used Swift/UVOT to observe 41P between May 7 – 9, 2017 and measured all the light within 1,600 km of the nucleus, including molecular emissions and sunlight reflected by dust grains. The light contributed by the small nucleus was negligible at this time, indicating that variations in brightness were dominated by the material recently released from the nucleus in our photometric aperture. The photometric variations are small and slowly varying (Fig. 2; Extended Data Table 2). Although the lightcurve is incomplete, the unobserved parts can reasonably be inferred, resulting in a single-peaked sinusoid (the hallmark of activity being modulated by changes in illumination induced by rotation) with a periodicity between 46 and 60 hours. The 14-hr range arises because the alignment of the overlapping segment of the phased sine curve is affected by changes in the comet's activity with its increasing distance to the Sun. We therefore conclude that during the two months of our observations, there was a significant change in the rotation period with an average increase between 0.40 – 0.67 hours per day. For discussions in this paper, we adopt 53 hr, the middle of our range, as our representative period.

[1] Department of Astronomy, University of Maryland, College Park, MD 20742, USA.

A CN morphology similar to the one seen in our DCT observations was observed between March 18 – 27 (12), but they report that the structure took 24 hours to repeat on March 19 and 21, and increased continuously to near 27 hours on March 26 and 27 (Fig. 3). During their densest coverage in late March, the morphology repeated at progressively later times on subsequent nights, revealing a daily trend that is consistent with our ensemble dataset from March to May. The consistent repetition of the morphology at the end of each lengthening period over such an extended time suggests that any non-principal axis (NPA) component of rotation is small. Furthermore, we cannot conceive a scenario where NPA rotation could mimic the observed continuously changing periodicity. Therefore, we assume that the nucleus is in a state of simple rotation.

Rotation periods have been measured for scores of comets, many with extensive coverage, but 41P is only the eighth comet for which a conclusive change in period has been measured, and both the fractional change and the rate of change far exceed those observed in the other comets (see Extended Data Table 3). Changes in comet rotation periods depend on the nucleus' size, shape, internal structure, activity, and rotational state (1, 2 4, 5). Comet 41P's nucleus radius is between 0.7 – 1.0 km (13), smaller than 70 to 90% of all measured radii of Jupiter-family comets (14). Its water production rate peaked around 2 x $10^{29}$ molecules/s in 2001 and 2 x $10^{28}$ molecules/s in 2006 (15). Our Swift observation suggests that production rates in 2017 were similar to those in 2006 (Extended Data Fig. 1). This implies that more than 50% of the comet's surface could be active, whereas most comets have less than 3% surface activity (16). Finally, while the comet's 20 hr rotation period in March was long compared to most comets, the >46 hr rotation period measured in May is among the longest known comet rotation periods (13). It is this combination of a slow rotation, high activity, and a small nucleus that contribute to the rapid changes of the rotation state of 41P.

However, these characteristics are not unique to 41P. In 2010, 103P/Hartley 2 had an initial period of 16.5 hours, a peak water production rate three times higher than 41P, and a smaller effective radius of 0.57 km (10). Even with the more extreme combination of these characteristics, its primary rotation period only changed by 2 hours in the 3 months around perihelion (17; Extended Data Table 3), more than an order of magnitude less than that of 41P. Thus, other factors must also play a role in producing a net torque that is much more efficient in 41P than in any other known comet. The Deep Impact fly-by of comet 103P allowed a close examination of the activity of its nucleus (10), and these details allow us to explore possible differences between it and 41P. The visible jets from Hartley 2 are primarily along the long axis, with little moment arm for producing torques; some of the water from 103P comes from icy grains in the coma, reducing the amount of gas contributing to torques (18, 19); and finally, the non-principal axis rotation of 103P acts to randomise the direction of the torques, reducing their efficiency.

Using the results from the four then-available comets that exhibited period changes an empirical parameter $X$ has been suggested that relates comet activity and changes in angular momentum (18). It was found that this parameter was nearly constant within a range of 1 – 2, leading to the conclusion that net torques are nearly the same irrespective of the effective active fractions of the nucleus. From our observations of comet 41P, we compute an $X$-parameter of over 50, inconsistent with that conclusion. ($X$-values for comets 19P/Borrelly (20) and 67P/Churyumov-Gerasimenko (21) also lie well outside the 1 – 2 range; Extended Data Table 3.) The deviation from this range implies that the torques, when integrated over all active areas, do not necessarily cancel out, and that the physical characteristics of nuclei greatly affect the torques' net efficiency. The effects of non-uniform activity and local topography are well illustrated by the results of the Rosetta mission to comet 67P/Churyumov-Gerasimenko where the rotation period first increased, then decreased as new parts of the comet's surface became illuminated (2). The active regions on the surface of 41P are likely oriented in a way such that its torques are highly optimised in comparison to many other comets.

We extrapolated the comet's rotation period to investigate its possible past and future behaviour (Fig. 4). Our model assumes that activity levels and effective torques were constant from apparition to apparition, e.g., that the orientation of the spin axis and water production did not change significantly. Our empirical model suggests that in the near future, the rotation period could exceed 100 hours. At such slow rotation rates the stabilising gyroscope effect disappears, and off-axis torques can tip the nucleus into an excited rotation state. If strong torques persist, then the nucleus can begin to spin up again with a different orientation of its rotational angular momentum vector. Such behaviour is consistent with simulations of the long-term evolution of spin states of small comet nuclei indicate that most comets go through a large change in their rotation period soon after their activation (5). This will lead to a temporary excitation of the nucleus' spin state, and for most comets the rotation period will slowly evolve afterwards. The simulations also show that in some cases, uniformly active surfaces can cause comets to respond unpredictably to changes in their spin state, and such comets may have inherently variable spin states, experiencing large changes in their rotation period during each perihelion passage.

Projecting back in time, 41P may have been near the critical fragmentation limit (around 5 hours; (3)) in the recent past. It is notable that the comet exhibited large outbursts in activity in 1973 and 2001 (15, 22), and these events may be related to its spin state evolution. The rotation may have caused the outbursts via spin-up fragmentation or landslides (23). Alternatively, the outbursts may have given rise to the spin evolution by exposing new active areas that generate outgassing torques.

## References

1. Jewitt, D. Cometary rotation: An overview. *Earth, Moon, and Planets* **79,** 35–53 (1997).
2. Keller, H. U., Mottola, S., Skorov, Y. & Jorda, L. The changing rotation period of comet 67P/Churyumov-Gerasimenko controlled by its activity. *Astron. Astroph*. **579,** L5 (2015).
3. Davidsson, B. J. R. Tidal Splitting and Rotational Breakup of Solid Biaxial Ellipsoids. *Icarus* **149,** 375–383 (2001).
4. Samarasinha, N. H., Mueller, B. E. A., Belton, M. J. S. & Jorda, L. Rotation of cometary nuclei. In *Comets II,* (eds. Festou, M., Keller, H. U., & Weaver, H. A.), 281–299 (University of Arizona Press, 2004).
5. Gutierrez, P. J., Jorda, L., Ortiz, J. L. & Rodrigo, R. Long-term simulations of the rotational state of small irregular cometary nuclei. *Astron. Astroph.* **406,** 1123–1133 (2003).
6. Gehrels, N. et al. The Swift Gamma-Ray Burst Mission. *ApJ* **611**, 1005–1020 (2004).
7. Farnham, T. L., Schleicher, D. G. & A'Hearn, M. F. The HB Narrowband Comet Filters: Standard Stars and Calibrations. *Icarus* **147,** 180–204 (2000).
8. A'Hearn, M. F., *et al.* Cyanogen jets in comet Halley. *Nature*, **324**, 649–651 (1986).
9. Schleicher, D. G., and Farnham, T. L., Photometry and imaging of the coma with narrowband filters. In *Comets II,* (eds. Festou, M., Keller, H. U., & Weaver, H. A.), 449–469 (University of Arizona Press, 2004).
10. A'Hearn, M. F. *et al.* EPOXI at Comet Hartley 2. *Science* **332,** 6036, 1396–1400 (2011).
11. Farnham, T. L., et al. Comet 41P/Tuttle-Giacobini-Kresák. CBET 4375 (2017).
12. Knight, M.M., Eisner, N., Schleicher, D.G., & Thirouin, A. Comet 41P/Tuttle-Giacobini-Kresák. CBET 4377 (2017).
13. Lamy, P. L., Toth, I., Fernandez, Y. R. & Weaver, H. A. in *Comets II* (eds. Festou, M., Keller, H. U., & Weaver, H. A.), 223–264 (University of Arizona Press, 2004).
14. Fernandez, Y. R. *et al.* Thermal properties, sizes, and size distribution of Jupiter-family cometary nuclei. *Icarus* **226,** 1138–1170 (2013).
15. Combi, M., SOHO SWAN Derived cometary water production rates collection, urn:nasa:pds:soho:swan_derived::1.0, (ed. L. Feaga), NASA Planetary Data System (2017).
16. A'Hearn, M. F., Millis, R. L., Schleicher, D. G., Osip, D. J. & Birch, P. V. The ensemble properties of comets: Results from narrowband photometry of 85 comets, 1976-1992. *Icarus* **118,** 223–270 (1995).
17. Knight, M. M. & Schleicher, D. G. CN morphology studies of comet 103P/Hartley 2. *Astron. J.* **141,** id. 183 (2011).
18. Samarasinha, N. H. & Mueller, B. E. A. Relating changes in cometary rotation to activity: current status and applications to comet C/2012 S1 (ISON). *Astroph. J. Lett.* **775,** L10 (2013).

19. Belton, M. J. Cometary evolution and cryovolcanism. *Canadian J. Physics* **90,** 807–815 (2012).
20. Mueller, B. E. A., and Samarasinha, N. H. Further investigation of changes in cometary rotation. *Asteroids, Comets, Meteoroids 2017* meeting, Montevideo, Uruguay. http://acm2017.uy/abstracts/Poster1.e.43.pdf (2017).
21. Mottola, S. *et al.* The rotation state of 67P/Churyumov-Gerasimenko from approach observations with the OSIRIS cameras on Rosetta. *Astron. Astroph.* **569,** L2 (2014).
22. Kronk, G. W., Catalog of Periodic Comets, http://cometography.com/pcomets/041p.html (2017).
23. Steckloff, J. K., Graves, K., Hirabayashi, T., Melosh, H. J. & Richardson, J. Rotationally induced surface slope-Instabilities and the activation of CO2 activity on comet 103P/Hartley 2. *Icarus* **272,** 60–69 (2016).

**Acknowledgements** We thank Mike Siegel and the Swift team for the planning of the observations of 41P. This research was supported by Swift Guest Investigator Program grant 1316125. We thank Audrey Thirouin, Chad Trujillo, and Nick Moskovitz for observing and/or donating telescope time to acquire images used to determine rotation periods from morphology. We thank Nora Eisner and Dave Schleicher for sharing their preliminary results with us. We thank Nalin Samarasinha for calculating the $\zeta$ parameter for 41P and 67P. These results made use of the Discovery Channel Telescope at Lowell Observatory. Lowell is a private, non-profit institution dedicated to astrophysical research and public appreciation of astronomy and operates the DCT in partnership with Boston University, the University of Maryland, the University of Toledo, Northern Arizona University and Yale University. The Large Monolithic Imager was built by Lowell Observatory using funds provided by the National Science Foundation (AST-1005313). This research has made use of NASA's Astrophysics Data System and of the JPL/Horizons ephemerides service, maintained by the JPL Solar System Dynamics group.

**Author contributions** DB and TLF designed and analysed the Swift observations. DB, TLF, and MSPK planned and acquired the DCT observations. TLF processed and analysed the DCT data. MSPK and DB modelled the period change. All authors wrote the manuscript.

**Author information** Correspondence and requests for materials should be addressed to dennis@astro.umd.edu. The authors declare no competing financial interests.

## Methods

**Photometry:** Swift/UVOT observations were obtained with the V-band filter, centred at 547 nm with a FWHM of 75 nm. We measured the brightness of the coma using photometry extracted from a circular aperture centred on the nucleus with a 1,600 km (10 – 11 arcsecond) radius at the distance of the comet. The median background flux was determined from an annulus with an inner radius of 50 arcseconds and an outer radius of 100 arcseconds (beyond the visible extent of the coma. We followed our standard calibration procedure (24) to derive the apparent magnitudes, *V.* These were then converted into absolute magnitudes, *H,* at 1 au to account for changes in the comet's geocentric distance $\Delta$, heliocentric distance $r_h$, and phase angle (using a phase function normalised to a phase angle of 90 degrees (25)) during our observation using the relation: $H = V - 5*\log \Delta - 5*\log r_h - 2.5*\log(P/P(90))$. The relation between the comet's activity and the heliocentric distance, which increased from 1.099 au to 1.108 au during the Swift observations, is currently not well constrained. This implies that a range of scale factors, $A$, are possible for the activity-corrected brightness $H'$ of the comet: $H' = H - A*\log(r_h/r_0)$. Larger scale factors imply longer rotation periods. We considered scale factors of $A = 0$ (i.e. an $r_h^2$ relation), $A = 28$ (an early empirical fit to the current brightness trend (26), and an upper limit of $A = 35$ (derived from an empirical fit to the brightness trend during the apparitions of 1995 and 2001 (26). As is shown in Extended Data Fig. 2, this results in a range of possible periods of repetition between 46 and 60 hours, with a central solution around 53 ± 0.5 hours ($A = 17$). Independent of the $r_h$ correction, periods shorter than 46 hours are not possible with our lightcurve (under our assumptions of simple rotation, and no outburst or other unusual activity).

There are too few measurements with the DCT to construct a meaningful lightcurve, and the night of Mar. 8 was not photometric (Cirrus clouds), which is why our observations focused on morphology rather than absolute measurements.

**Production rates:** We used Swift/UVOT images to determine water production rates following the method outlined by Bodewits et al. (24). The UVW1 filter (central wavelength 260 nm, FWHM 70 nm) encompasses the three strong OH A-X transitions. We first created stacks containing all UVW1 images and V-band images acquired between May 4 – 9, 2017 using a clipped mean routine. We then removed the continuum contribution to the light measured with this filter by subtracting a weighted V-band image. There was no obvious repetitive morphology in the OH images. Fluxes in apertures with radii between 5 and 300 arcsec (775 to 46,500 km at the comet) were converted into OH column densities assuming fluorescence rates at the heliocentric velocity and distance of the comet (27). Production rates were derived using a vectorial model (28).

**Active area:** We derived the minimum active area corresponding to the measured water production rate using a sublimation model that assumes that every surface element has constant solar elevation (as would be the case if the spin axis was pointed at the Sun, i.e. an obliquity of 90 degrees, or if the nucleus was very slowly rotating) and is therefore in local, instantaneous equilibrium with sunlight (29). This maximises the sublimation averaged over the entire surface and results in a minimum total active area. We further assumed a Bond albedo of 0.02 and 100% infrared emissivity. The modelled $H_2O$ sublimation rate is 3.35 x $10^{17}$ molecules/cm$^2$ at 1.05 au. Assuming a peak water production rate of 2 x $10^{28}$ molecules/s (Extended Data Fig. 1) we find an active area of at least 6 km$^2$, equivalent to an active fraction of 50-97% of a spherical nucleus with a radius between 0.7 – 1 km.

**Modelling the change in rotation period:** To extrapolate the rotation period of 41P to past and future apparitions we used the relation between the rate of change of the angular velocity $d\omega/dt$, the comet's water mass loss rate $Q$, and the radius of the nucleus $R$ (18):

$$d\omega/dt = C * Q(t) / R^4 \qquad (1)$$

We assumed a nucleus with a radius of 0.7 km and used our measurements of the production rate and the average change of rotation period during the current apparition to empirically determine the constant $C$. To estimate the orbital gas mass loss, we used the empirical relation between the comet's brightness and the heliocentric distance $Q \sim r_h^{-4.8}$ and fitted this to the SOHO measurements of water production rates during the comet's 2006 apparition (15). We assumed abundances of 10% for both CO and $CO_2$ relative to water, and that activity beyond 3 au is negligible. When the nucleus reaches a rotation frequency of 0, the period is infinite, hence the growth off the top of the figure. At this point in the model, the rotation reverses (rotational pole flip) and the period decreases from infinity. However, in reality the rotation will become excited, the illumination on the surface will change, and the torques should also change.

Integrating the gas production rates from 3 au before to 3 au after perihelion results in a mass-loss rate of 6 x $10^9$ kg in volatiles per orbit, or about 1% of the nucleus mass for a density of 500 kg/m$^3$.

The models of Gutierrez et al. (5) assume a certain initial spin state and their evolution is modeled for 10 to 100 orbits. Comet 41P has orbited the Sun approximately 30 times since its discovery in 1858. Their paper shows several scenarios that settle on hyperbolic evolution after ~10 to 30 orbits, and comets whose spin states keep evolving throughout the simulations. However, as pointed out by Samarasinha et al (4), the Gutierrez models did not explore the full parameter space and we are hesitant to imply a more quantitative interpretation of the models.

**Data Availability:** All Swift/UVOT data is available from the Barbara A. Mikulski Archive for Space Telescopes (https://archive.stsci.edu) and from the Swift Archive Portal (http://www.swift.ac.uk/swift_portal/) under program ID 1316125. The photometric measurements are provided as a source file for Fig. 2. Other data that support the findings of this study are available from the corresponding author upon reasonable request.

## References

24. Bodewits, D. *et al.* The evolving activity of the dynamically young comet C/2009 P1 (Garradd). *Astroph. J.* **786,** 48–57 (2014).
25. Schleicher, D. Composite dust phase function for comets, http://asteroid.lowell.edu/comet/dustphase.html (2017)
26. Yoshida, S. Seiichi Yoshida's home page. http://www.aerith.net/comet/catalog/0041P/2017.html (2017).
27. Schleicher, D. G. & A'Hearn, M. F. The fluorescence of cometary OH. *Astroph. J.* **331,** 1058–1077 (1988).
28. Festou, M. C. The density distribution of neutral compounds in cometary atmospheres. I - Models and equations. *Astron. Astroph.* **95,** 69–79 (1981).
29. Cowan, J. J. & A'Hearn, M. F. Vaporization of comet nuclei - Light curves and life times. *Moon and the Planets* **21,** 155–171 (1979).
30. A'Hearn, M. F. *et al.* Deep Impact: Excavating Comet Tempel 1. *Science* **310,** 258–264 (2005).
31. Knight, M.M., et al. A quarter-century of observations of comet 10P Tempel 2 at Lowell Observatory: continued spin-down, coma morphology, production rates, and numerical modeling. *Astroph. J.* **144**, id. 153 (2012).
32. Mueller, B. E. A. and Ferrin, I. Change in the rotational period of comet P/Tempel 2 between the 1988 and 1994 Apparitions. *Icarus* **123**, 2, 463-477.
33. Mueller, B. E. A, Samarasinha, N. H., Rauer, H., and Helbert, J. Determination of a precise rotation period for the Deep Space 1 target, comet 19P/Borelly. *Icarus*, **209**, 2, 745-752 (2010).
34. Sierks, H. *et al.* On the nucleus structure and activity of comet 67P/Churyumov-Gerasimenko. *Science* **347,** aaa1044 (2015).
35. ESA Flight Dynamics Team, Comet Rotation Period, http://sci.esa.int/rosetta/58367-comet-rotation-period/ (2017).
36. Schleicher, D. G., Millis, R. L., Osip, D. J. Comet Levy (1990c): Ground-based photometric results. *Icarus* **94**, 2, 511–523 (1991).
37. Feldman, P. D., Budzien, S. A., Festou, M. C., A'Hearn, M. F. & Tozzi, G. P. Ultraviolet and visible variability of the coma of Comet Levy (1990c). *Icarus* **95,** 65–72 (1992).

## Figures

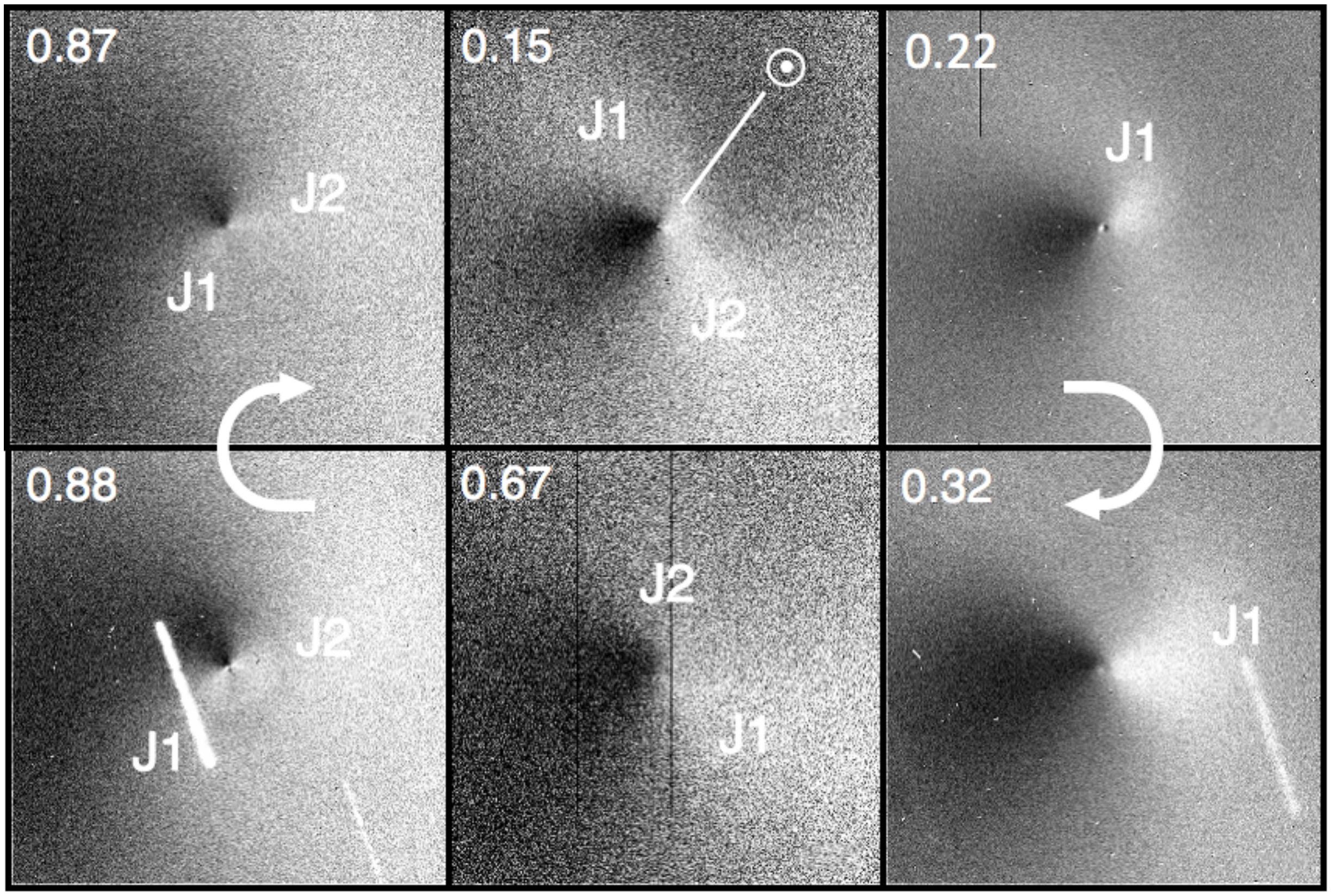

**Figure 1 | Repeating CN jets in the coma of comet 41P/Tuttle-Giacobini-Kresák.** Sequence of DCT images showing the cyanogen coma of comet 41P, enhanced to reveal two rotating jets (labelled J1 and J2). The images, obtained on 7 and 8 March 2017, progress in a clockwise direction, as indicated by the curved arrows. Nearly identical morphologies are seen in the left two panels, which were obtained 20.1 h apart, and the sequence suggests these two images are slightly more than one full rotation apart, leading to the derived 19.9 h period. The other panels, labelled in the upper left corner with the fraction of the period (phase) when the image was obtained, show a continuously changing morphology that precludes any periods that are sub-multiples of the 19.9-h derived value. Each panel spans approximately 20,000 km at the distance of the comet, is centred on the position of the nucleus (too small to be resolved), and is oriented with north up and east to the left. The direction to the Sun is indicated by ⊙. Images were enhanced by dividing out the averaged azimuthal profile. Regions that are brighter than average at that distance from the nucleus are white while regions that are fainter are black. The white streaks are background stars.

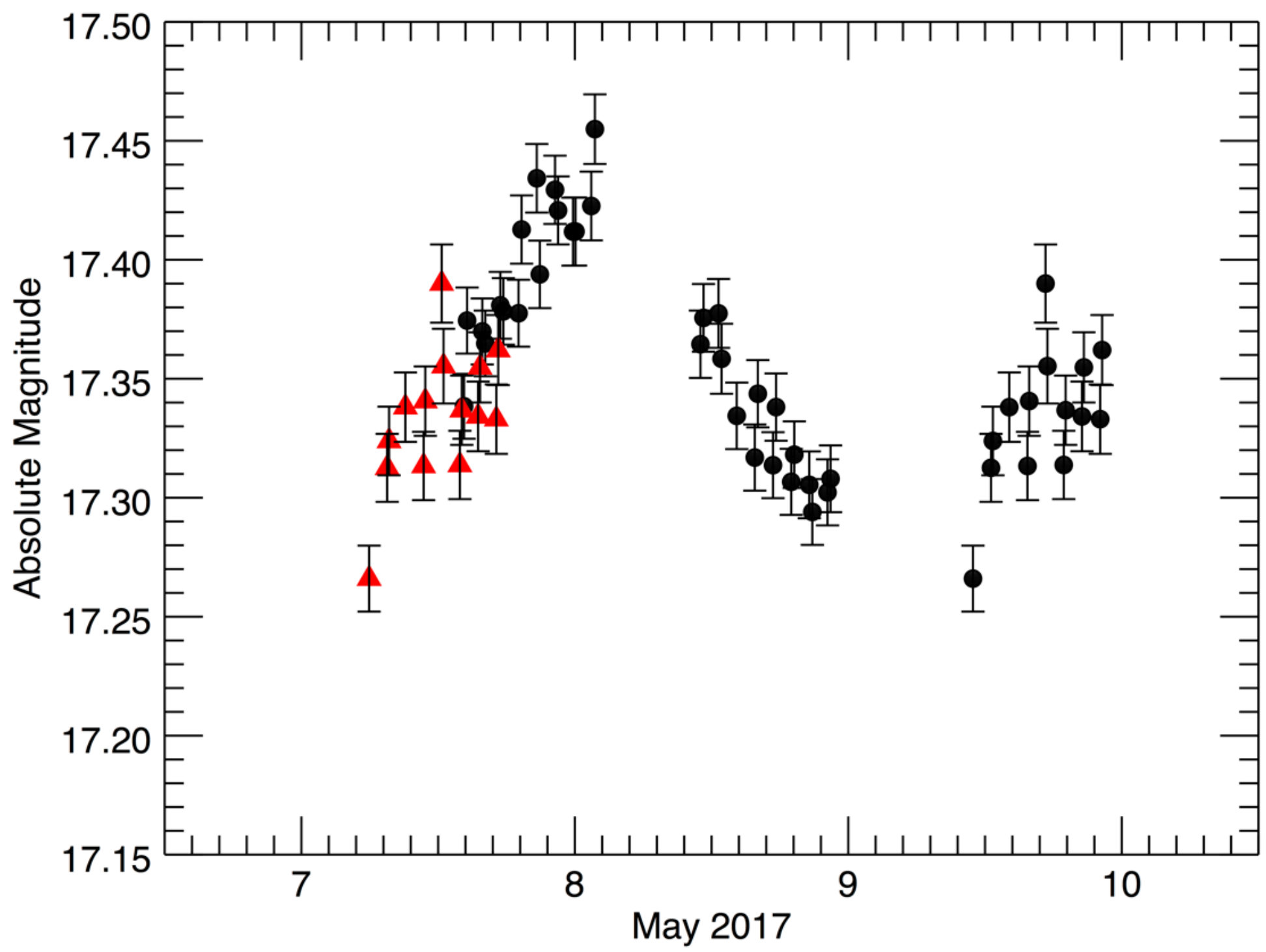

**Figure 2 | Inner coma light curve measured by Swift/UVOT between May 7 – 10, 2017.** The data acquired between May 9.4 – 10 are repeated as red triangles (▲), phase-shifted to best match the data acquired between May 7.5 – 8 (●). Depending on the decrease of the comet's activity with the heliocentric distance (Methods), a range of periods between 46 – 60 hours is found (see Extended Data Fig. 2). The central period, 53 hours, is shown here. Error bars indicate 1-sigma stochastic uncertainties.

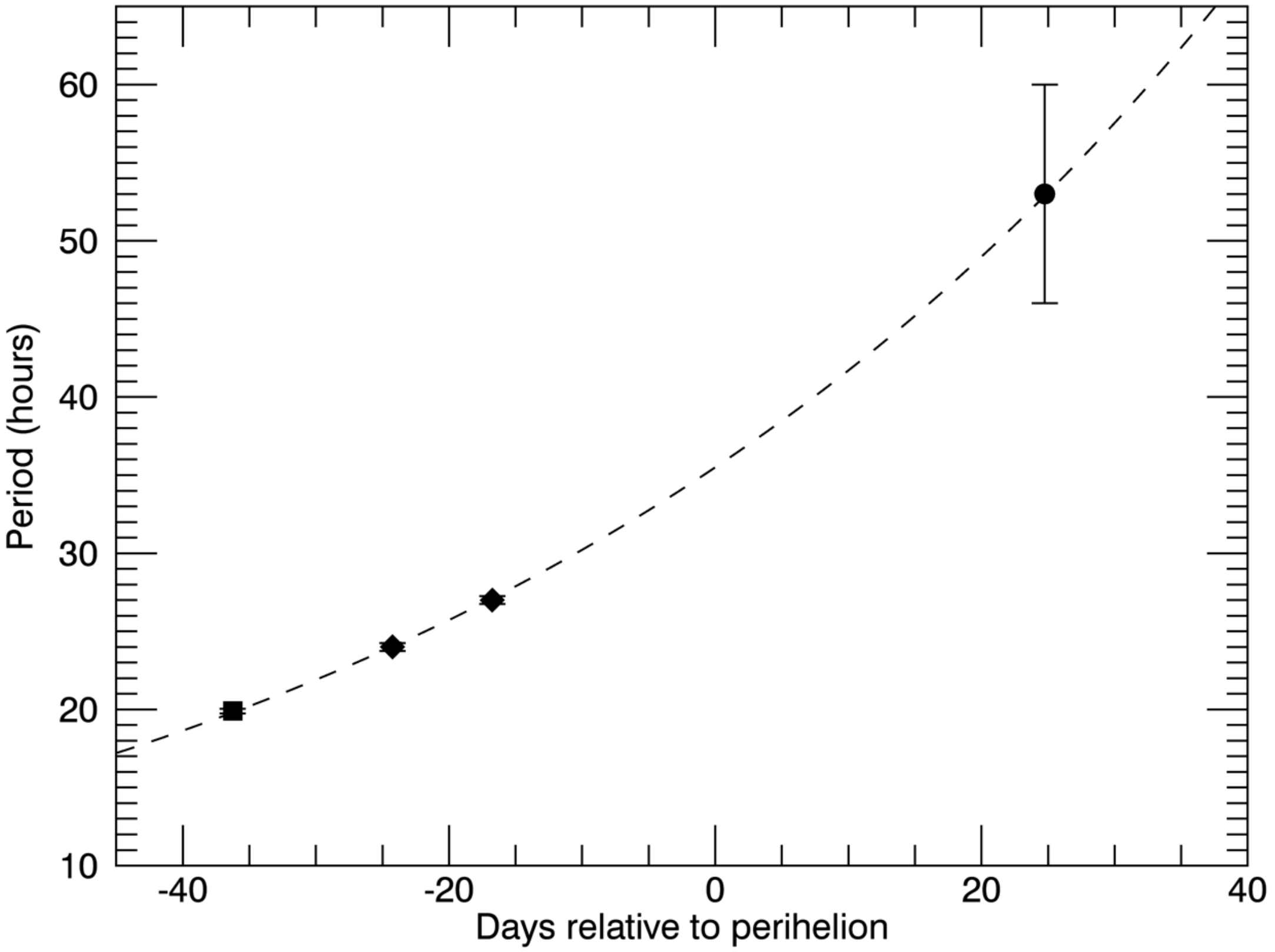

**Figure 3 | Rotation periods for 41P measured as a function of time relative to perihelion.** Perihelion occurred on April 12, 2015. The period increased at an average rate of 0.53 hours/day over more than 60 days, an unprecedented rate of change. The different observations are indicated by symbols: Swift (●; this work); DCT (■; 11 and this work); and results acquired by another team using Lowell's 31" telescope (◆; 12). The dashed line is drawn to guide the eye. Error bars indicate 1-sigma absolute uncertainties; the error bar on the last point indicates the range of possible solutions for the Swift measurements due to the uncertainty in the change of activity as a function of the heliocentric distance.

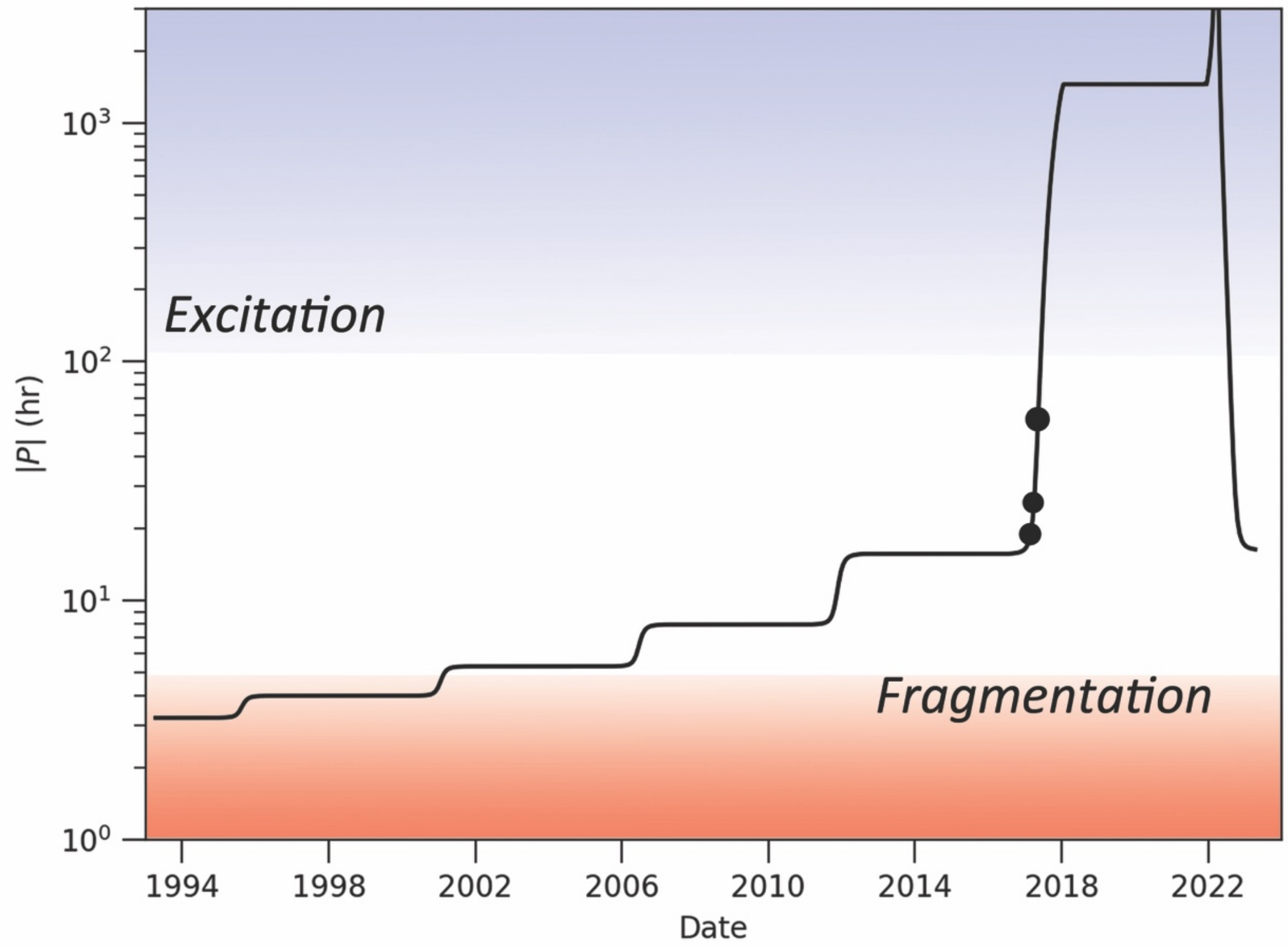

**Figure 4 | Extrapolation of the rotation period of 41P in time.** We modelled past and future absolute values of the period by extrapolating the 2017 torques to other perihelion passages (upward steps). This scenario suggests that the nucleus could evolve from rapid rotation near the fragmentation limit (3) to an excited, unstable spin state (4,5) in only a few orbits. The 2017 observations are indicated by filled dots (●).

**Extended Data Table 1 | Summary of measure rotation periods.** *Δ* is the geocentric distance of the comet, $r_h$ denotes its heliocentric distance.

| Telescope | Dates | *Δ* | $r_h$ | Rotation Period | References |
|---|---|---|---|---|---|
| | (UTC) | (au) | (au) | (hours) | |
| DCT/LMI | Mar. 6 – Mar. 9, 2017 | 0.24 – 0.18 | 1.22 - 1.16 | 19.9 ± 0.15 | This work; 11 |
| Lowell 31” | Mar. 18 – 27, 2017 | 0.16 - 0.14 | 1.1 - 1.06 | 24 - 27 ± 0.25 | (12) |
| Swift/UVOT | May 6 – 9, 2017 | 0.21 | 1.1 | 46 – 60 | This work |

**Extended Data Table 2 | Observing log of Lowell Observatory's Discovery Channel Telescope.** *Δ* is the geocentric distance of the comet, $r_h$ denotes its heliocentric distance.

| Date (UTC) | Midtime (UTC) | $r_h$ (au) | *Δ* (au) | Phase (deg.) | Rot. Phase | Observers |
|---|---|---|---|---|---|---|
| Mar. 6, 2017 | 2:38 | 1.16 | 0.20 | 27.5 | 0.32 | Thirouin/Moskovitz |
| Mar. 7, 2017 | 5:27 | 1.16 | 0.19 | 29.0 | 0.67 | Farnham/Kelley/Bodewits |
| Mar. 7, 2017 | 9:33 | 1.15 | 0.19 | 29.3 | 0.87 | Farnham/Kelley/Bodewits |
| Mar. 8, 2017 | 3:41 | 1.15 | 0.19 | 30.3 | 0.78 | Farnham/Kelley/Bodewits |
| Mar. 8, 2017 | 5:40 | 1.15 | 0.19 | 30.4 | 0.88 | Farnham/Kelley/Bodewits |
| Mar. 8, 2017 | 8:23 | 1.15 | 0.19 | 30.6 | 0.02 | Farnham/Kelley/Bodewits |
| Mar. 8, 2017 | 11:04 | 1.15 | 0.19 | 30.7 | 0.15 | Farnham/Kelley/Bodewits |
| Mar. 9, 2017 | 8:15 | 1.14 | 0.18 | 32.0 | 0.22 | Thirouin/Moskovitz |

**Extended Data Table 3 | Characteristics of other comets for which a change in rotation period has been measured.** *) Interval between rotation period measurements, which may not reflect the time it took to change. In some instances, period changes have been observed on multiple orbits. **) For 67P/Churyumov-Gerasimenko characteristics before and after perihelion have been given separately.

| **Name** | ***P* (hr)** | **$\Delta P$ (hr)** | **Interval* (yr)** | **$\Delta P$/orbit (hr/orbit)** | **Effective Radius (km)** | **Active Fraction (%)** | ***X*** | **References** |
|---|---|---|---|---|---|---|---|---|
| 41P/T-G-K | 20 | >26 | 0.2 | >26 | <1 | >50 | 78 | This paper; 13 |
| 2P/Encke | 11 | -0.072 | 3.3 | -0.072 | 2.4 | 1.0 | 1.0 | 13, 16, 18 |
| 9P/Tempel 1 | 41 | -0.233 | 5.52 | -0.233 | 2.8 | 6.5 | 1.1 | 16, 18, 30 |
| 10P/Tempel 2 | 9 | 0.018 | 22.5 | 0.045 | 5.3 | 0.8 | 1.9 | 13, 16, 31, 32 |
| 19P/Borelly | 29 | >0.667 | 13 | 0.33 | 2.4 | 9.1 | 5.1 | 16, 20, 31, 33 |
| 67P/C-G (pre) **) | 12 | 0.027 | 1.21 | 0.027 | 1.65 | 2.0 | 0.4 | 2, 21, 34, 35 |
| 67P/C-G (post) **) | | -0.375 | 1.33 | -0.375 | | | 6.1 | 2, 21, 34, 35 |
| 103P/Hartley 2 | 17 | 2 | 0.25 | 2 | 0.57 | >100 | 1.5 | 13, 17, 18, 19 |
| Levy (1990c) | 19 | -1.3 | 0.058 | -1.3 | - | - | - | 36, 37 |

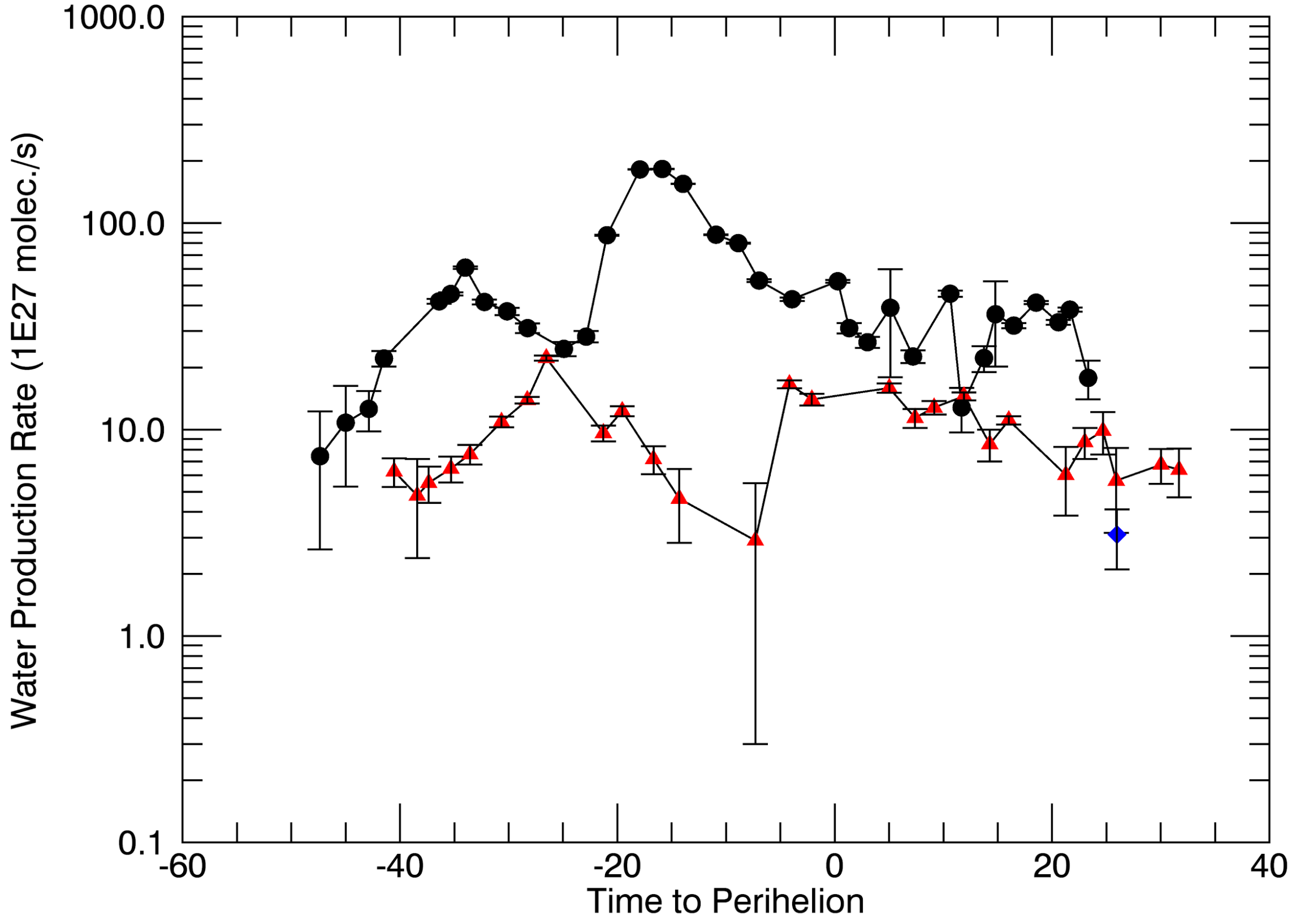

**Extended Data Figure 1 | Water production rates of 41P/Tuttle-Giacobini-Kresák in 2001, 2006, and 2017.** Production rates were derived from Hydrogen Lyman-α emission observed by the SWAN instrument on board the SOHO spacecraft (15) in 2001 (●) and 2006 (▲). For the SWAN data, 1-sigma stochastic errors a shown; systematic uncertainties are at the 30% level (15). We used Swift/UVOT observations of Hydroxyl (OH) emission to determine the water production rate in 2017 (◆). For the Swift data, the error bars represent the systematic uncertainty. The comet had two 4-magnitude outbursts in optical wavelengths just before its perihelion in 2001 (22), and these correspond to the peaks at approximately 35 and 15 days before perihelion.

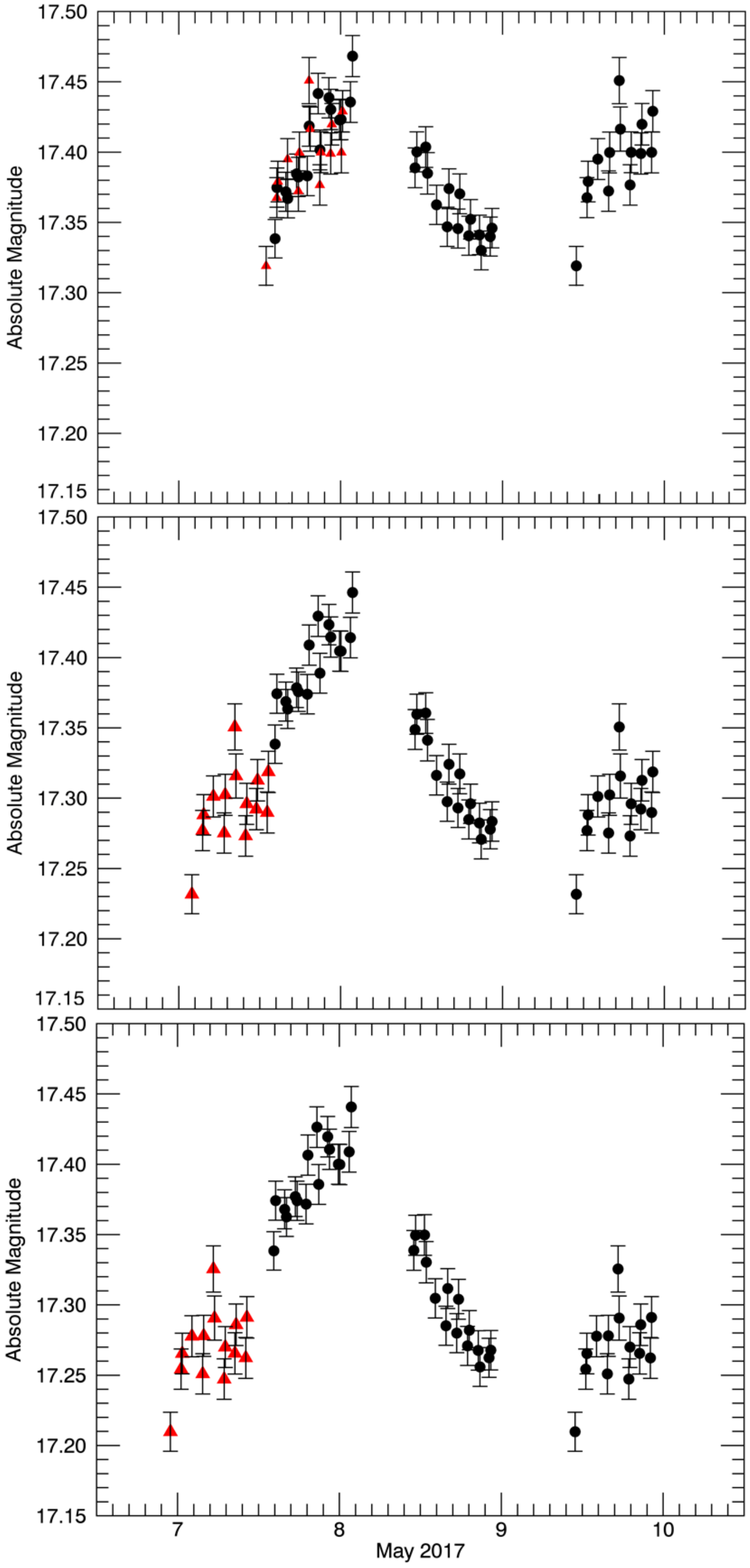

**Extended Data Figure 2 | Rotation period solutions for different activity models.** Absolute magnitudes based on Swift/UVOT photometry (black dots ●) corrected for different

relationships ($A$) between the comet's activity and its distance to the Sun (See Methods). An increase in ($A$) corresponds to an increase in the rotation period needed to phase the overlapping sine curve segment (red triangles ▲). Panel a: A = 0, period = 46 hours. Panel b: $A = 28$, period = 57 hours. Panel c: $A = 35$, period = 60 hours. Error bars indicate 1-sigma stochastic uncertainties.